\documentstyle[pre,aps,amsfonts,amssymb,twocolumn,eqsecnum,psfig]{revtex}
\begin{document}
\def\u{\bbox} 
\def\mathcal#1{{\cal #1}}
\def\phi{\varphi}     
\def\epsilon{\varepsilon}
\def\d{\displaystyle}
\def\o{\omega}
\def\I{{\Bbb I}^-}
\def\Pa{\u{\partial}}

\draft
\title{Strange attractor for the renormalization flow for invariant tori of
Hamiltonian systems with two generic frequencies}
\author{C.\ Chandre and H.R.\ Jauslin}
\address{Laboratoire de Physique-CNRS, Universit\'e de Bourgogne,
B.P.\ 47 870, F-21078 Dijon, France}
\maketitle

\begin{abstract}
We analyze the stability of invariant tori for Hamiltonian systems
with two degrees of freedom by constructing a transformation that
combines Kolmogorov-Arnold-Moser theory and renormalization-group techniques.
This transformation is based on the continued fraction expansion 
of the frequency of the torus.
We apply this transformation numerically for arbitrary frequencies
that contain bounded entries
in the continued fraction expansion. 
We give a global picture of renormalization flow for the stability of
invariant tori, and we
show that the properties of critical (and near critical) tori 
can be obtained by analyzing renormalization dynamics around 
a single hyperbolic strange attractor. We compute the fractal diagram, i.e.\
the critical coupling as a function of the frequencies,
associated with a given one-parameter family.
\end{abstract}


\section{Introduction}

Invariant tori play a fundamental role in the stability of Hamiltonian systems.
For two degrees of freedom, they act as barriers in phase space.
Renormalization-group (RG) transformations have been proposed to understand 
the breakup of invariant tori for area-preserving maps~\cite{mack83,Bmack93} and
for Hamiltonian systems with two degrees
of freedom~\cite{gall87,koch99,govi97,chan98a,chan98b,abad98}. 
The RG-transformations described in 
Refs.~\cite{koch99,chan98b,abad98} combine an elimination of the irrelevant 
part of the perturbation, and a rescaling of phase space which is adapted
to the frequency of the specific invariant torus.
These transformations are constructed as canonical changes of
coordinates.
They have been implemented numerically 
for the golden mean torus (with frequency $\gamma=(\sqrt{5}-1)/2$). Based on
the numerical analysis of the renormalization flow, the following picture
emerges :
A trivial fixed point $H_0$
(which is an integrable Hamiltonian)
of the RG-transformation characterizes the domain where the
Hamiltonians have a smooth invariant torus with frequency $\gamma$, i.e., all
Hamiltonians attracted to $H_0$ by renormalization are locally canonically
conjugated to an integrable Hamiltonian. The boundary of the domain of attraction
of $H_0$ is the domain where the tori are critical, at the threshold of their
breakup. This boundary (named the {\em critical surface})
is expected to be of codimension 1, and to coincide with the set of Hamiltonians 
with critical coupling, beyond which they 
do no longer have an invariant torus with frequency $\gamma$.
The critical surface is expected to be the stable
manifold of a nontrivial fixed point $H_*$.  
This global picture is still at the stage of conjecture : In the
perturbative regime, it is supported by rigorous results~\cite{koch99}. From the
numerical analysis of the neighborhood of $H_*$, the scaling properties of
critical tori~\cite{kada81,shen82} are found by these renormalization
transformations : It is
a strong argument in favor of the link between $H_*$ and critical invariant tori.
Furthermore, this renormalization approach gives a method to calculate the
critical coupling for a given one-parameter family (the value of the parameter
for which the considered invariant torus is critical).
It has been verified that this critical coupling coincides with
the critical couplings obtained by other methods like Greene's criterion~\cite{gree79}
or Laskar's frequency analysis~\cite{lask93}.\\

Since the relevant quantity for stability is the ratio between the two components of the
frequency vector,
we consider an invariant torus with frequency vector $\u{\omega}_0=(\omega,-1)$
where the frequency $\omega$ is irrational (the frequency
vector must satisfy a Diophantine condition). The purpose of the RG-transformations
we define is to analyze the stability of this invariant torus for
Hamiltonians with two degrees of freedom. The idea is to set up
a transformation $\mathcal{R}$ as a canonical change of coordinates
that maps a Hamiltonian $H$ into a rescaled
one $\mathcal{R}(H)$, such that irrelevant degrees of freedom are eliminated.
The transformation $\mathcal{R}$ should have the following properties :
$\mathcal{R}$ has an attractive fixed set (trivial fixed set) of integrable
Hamiltonians. Every Hamiltonian in its domain of attraction $\mathcal{D}$ has
a smooth invariant torus with frequency vector $\u{\omega}_0$. The aim is to
show that there is another fixed set which lies on the boundary $\partial \mathcal{D}$
(the critical surface) and that is attractive for every Hamiltonian on 
$\partial \mathcal{D}$.\\
Renormalization-group approaches for the breakup of invariant tori for 
arbitrary frequency have been proposed for circle 
maps~\cite{ostl83,farm85,umbe86,lanf86,lanf88}
and for area-preserving maps~\cite{sati87}. The boundary of Siegel disks has also been
analyzed by renormalization~\cite{mack87}. The numerical results suggest that
{\em statistical} self-similarity characterizes the considered problems at criticality.
It has been conjectured that it can be described by an ergodic attractor of a
renormalization transformation.\\
The numerical implementation of the RG-transformation for Hamiltonian 
systems with two degrees of freedom, gives support
to this picture : We conjecture that the properties (scaling factors 
and critical exponents) of critical tori can be 
obtained by analyzing renormalization dynamics around a {\em single} strange chaotic attractor.
Each critical torus displays a sequence of critical exponents, which are
the eigenvalues of the linearized renormalization map along its trajectories.  
Two different critical tori display the same set of critical exponents
but with a different probability distribution. In that sense, we can speak
of a single universality class describing
critical invariant tori for Hamiltonian systems with two degrees of freedom.\\

In Sec.~\ref{sec:RG}, we construct the RG-transformations for
an invariant torus with bounded
entries in the continued fraction expansion of its frequency. In Sec.~\ref{sec:res},
we apply this construction numerically for a set of frequencies which have
only 1 and 2 in their continued fraction expansion. We show that a critical 
strange attractor can be expected to describe the properties of critical invariant tori.
In Sec.~\ref{sec:frac}, we compute a fractal diagram which is the set of 
critical couplings $\varepsilon_c(\omega)$ for a given one-parameter family of
Hamiltonians.
In Sec.~\ref{sec:simple}, we construct a simple approximate RG-transformation
that gives qualitatively all the relevant features of the renormalization
dynamics.

\section{Renormalization transformation}
\label{sec:RG}

We describe the renormalization scheme for a torus with arbitrary
frequency vector $\u{\o}_0=(\o,-1)$ where $\o \in ]0,1[$.
This renormalization relies upon the continued fraction expansion of 
$\o$ :
$$ \o=\frac{1}{a_0+\d\frac{1}{a_1+\cdots}} \equiv [a_0,a_1,\ldots].$$
The best rational approximants of $\o$ are given by the truncations of this 
expansion : $p_k/q_k=[a_0,a_1,\ldots,a_k=\infty]$. The
corresponding periodic orbits with frequency vectors $\u{\omega}_{\u{\nu}_k}=(p_k/q_k,-1)$
accumulate at the invariant torus. We call ``resonance'', a vector $\u{\nu}_k=(q_k,p_k)$ 
(for $k\geq 1$) which is orthogonal to $\u{\omega}_{\u{\nu}_k}$.
The word resonance refers to the fact that the small denominators
$\u{\o}_0\cdot\u{\nu}_k$ that appear in the perturbation expansion
are the smallest ones, i.e., $|\u{\o}_0\cdot\u{\nu}_k|<|\u{\o}_0\cdot\u{\nu}|$
for any $\u{\nu}=(q,p)$ different from zero and $\u{\nu}_k$, and such that
$|q|<q_{k+1}$~\cite{Bcass57}.
The sequence of resonances satisfies
$|\u{\o}_0\cdot\u{\nu}_{k+1}| < |\u{\o}_0\cdot\u{\nu}_k|$ and
$
\lim_{k \to \infty} |\u{\o}_0\cdot\u{\nu}_k|=0,
$
as can be seen from the relation
\begin{equation}
\label{eqn:res}
\u{\nu}_k=N_{a_0}\cdots N_{a_{k-1}}\u{\nu}_0,
\end{equation}
where $\u{\nu}_0=(1,0)$ and $N_{a_i}$ denotes the matrix
$$ N_{a_i}=\left( \begin{array}{cc} a_i & 1 \\ 1 & 0 \end{array} \right).$$
Moreover, $\u{\o}_0\cdot\u{\nu}_k$ and $\u{\o}_0\cdot\u{\nu}_{k+1}$ are of
opposite sign (as the stable eigenvalue of $N_{a_i}$ is negative); thus
the torus is approached from above and from below by the sequence of periodic
orbits with frequency vectors $\{ \u{\omega}_{\u{\nu}_k} \}$.\\
 
The main scale of the torus
is characterized by the periodic orbit with frequency vector $\u{\omega}_{\u{\nu}_0}=(0,-1)$ for 
the torus with frequency vector $\u{\omega}_0=(\o,-1)$. The
next smaller scale is characterized by $\u{\omega}_{\u{\nu}_1}=(1/a_0,-1)$. 
The renormalization transformation, denoted $\mathcal{R}_{a_0}$,
changes the coordinates such that the next smaller scale becomes the main one,
i.e., the main scale in the new coordinates is characterized by 
$\u{\omega}_{\u{\nu}_0}'=\u{\omega}_{\u{\nu}_1}$. It acts as a microscope in phase
space : it looks the system at a smaller scale in phase space, and at a longer time scale.\\

We consider Hamiltonians $H$ written in terms of actions
$\u{A}=(A_1,A_2)\in {\Bbb R}^2$ and angles 
$\u{\varphi}=(\varphi_1,\varphi_2)\in{\Bbb T}^2$, of the form
\begin{equation}
\label{eqn:HRG}
H(\u{A},\u{\varphi})=H_0(\u{A})+V(\u{\Omega}\cdot\u{A},\u{\varphi}),
\end{equation}
where $\u{\Omega}=(1,\alpha)$ is a vector not parallel to the
frequency vector $\u{\o}_0$, and $H_0$ is given by 
$$
H_0(\u{A})=\u{\omega}_0\cdot\u{A}+\frac{1}{2}(\u{\Omega}\cdot\u{A})^2.
$$
We notice that the invariant torus with frequency vector $\u{\o}_0$ is located,
for $H_0$, at $\u{A}$ such that 
$\u{\Omega}\cdot\u{A}=0$ and $\u{\o}_0\cdot\u{A}=E$, where $E$ is the total energy of the
system.\\
The RG-transformation $\mathcal{R}_{a_0}$
consists of four steps :\\
{\bf (1)} A shift of the resonances constructed from the
condition $\u{\nu}_1 \mapsto \u{\nu}_0$ :
we require that $\cos[(a_0,1)\cdot \u{\phi}]=\cos [(1,0)\cdot\u{\phi}']$.
This change  
is done via a linear canonical transformation 
$$
(\u{A},\u{\phi})\mapsto (\u{A}',\u{\phi}')=
(N_{a_0}^{-1}\u{A},N_{a_0}\u{\phi}).
$$
This step changes the frequency vector $\u{\o}_0$ into $\u{\o}_0'=(\o',-1)$,
since $N_{a_0}\u{\o}_0=-\o\u{\o}_0'$, where the frequency $\o'$ is given by the
Gauss map
\begin{equation}
\label{eqn:gauss}
\omega \mapsto \omega'=\omega^{-1}-\left[ \omega^{-1}\right],
\end{equation}
and $\left[\omega^{-1} \right]$ denotes the integer part of $\omega^{-1}$.
Expressed in terms of
the continued fraction expansion of the frequency, it corresponds to a shift to the left
$$
\o=[a_0,a_1,a_2,\ldots]\mapsto \o'=[a_1,a_2,a_3,\ldots].
$$
The sequence of the resonances $\{\u{\nu}_k\}$ is mapped into the sequence
$$
\u{\nu}_k'=N_{a_1}\cdots N_{a_{k-1}}\u{\nu}_0.
$$
The linear term in the actions of $H_0$ is changed into $-\o \u{\omega}_0'\cdot\u{A}'$.\\
{\bf (2)} We rescale the energy 
by a factor $\o^{-1}$ (or equivalently time by a factor $\o$), and we change 
the sign of both phase space coordinates
$(\u{A},\u{\phi})\mapsto (-\u{A},-\u{\phi})$, in
order to have $\u{\o}_0'$ as the new frequency vector, i.e., the 
linear term in the actions of $H_0$ is $\u{\o}_0'\cdot\u{A}$. Furthermore,
$\u{\Omega}=(1,\alpha)$ is changed into $\u{\Omega}'=(1,\alpha')=(1,(a_0+\alpha)^{-1})$.
The map $\alpha\mapsto (a_0+\alpha)^{-1}$ is the inverse of the Gauss map, i.e.\
if $\alpha=[b_0,b_1,\ldots]$, then $\alpha'=[a_0,b_0,b_1,\ldots]$. We remark that
if $\alpha$ has the continued fraction expansion $\alpha=[b_0,b_1,\ldots]$, and if we
define the two-sided sequence
$$
[\alpha|\omega]=[\ldots,b_2,b_1,b_0|a_0,a_1,a_2\ldots],
$$
then the map $[\alpha|\omega]\mapsto [\alpha'|\omega']$ corresponds to the (two-sided)
Bernoulli shift
$$
[\ldots,b_2,b_1,b_0|a_0,a_1,a_2,\ldots] \mapsto [\ldots,b_2,b_1,b_0,a_0|a_1,a_2,\ldots].
$$
\noindent {\bf (3)} Then we perform a rescaling of the actions : $H$ is changed into
$$
H'(\u{A},\u{\phi})=\lambda H\left(\frac{\u{A}}{\lambda},\u{\phi}\right),$$
with 
$\lambda=\lambda(H)$ such that the mean value (i.e., the average over the angles) 
of the quadratic term in 
the actions in $H'$ is equal to $1/2$. This normalization
condition is essential for the convergence of the transformation. 
After Steps 1, 2 and 3, the Hamiltonian expressed
in the new variables is
\begin{equation}
H'(\u{A},\u{\varphi})=\lambda\omega^{-1} H\left(-\frac{1}{\lambda}N\u{A},
-N^{-1}\u{\varphi}\right).
\end{equation}
For $H$ given by Eq.\ (\ref{eqn:HRG}), this expression becomes
\begin{eqnarray}
H'(\u{A},\u{\varphi})=&&\u{\omega}_0'\cdot\u{A}+\frac{\omega^{-1}}{2\lambda}
(a_0+\alpha)^2 (\u{\Omega}'\cdot\u{A})^2\nonumber \\
&&+\lambda\omega^{-1} V\left(-\frac{a_0+\alpha}{\lambda}
\u{\Omega}'\cdot\u{A},-N^{-1}\u{\varphi}\right).
\end{eqnarray}
Thus the choice of the rescaling in the actions (Step 3) is
\begin{equation}
\label{eqn:rescaling}
\lambda=\omega^{-1} (a_0+\alpha)^2 (1+2\langle V^{(2)}\rangle),
\end{equation}
where $\langle V^{(2)}\rangle$ denotes the mean value 
of the quadratic part of $V$,
in the $(\u{\Omega}'\cdot\u{A})$-variable.\\
{\bf (4)} The last step is a canonical transformation that eliminates the nonresonant part of 
the perturbation in $H'$.\\
The choice of which part of the perturbation is to be considered resonant or not is somewhat 
arbitrary. The set of nonresonant modes includes the modes of the perturbation which
are sufficiently far from the resonances in order to avoid small denominator problems
during this elimination step.
A reasonable choice for the nonresonant modes
is the set
$I^-$ of integer vectors $\u{\nu}\in{\Bbb Z}^2$ such that $|\nu_2|> |\nu_1|$.
A mode which is not an element of $I^-$, will be called resonant.
We notice that Eq.~(\ref{eqn:res}) defining $\u{\nu}_k=(q_k,p_k)$ shows that
$q_k\geq p_k$ for $k\geq 0$, i.e., the resonances are not elements of
$I^-$.
From the form of the eigenvectors of $N_{a_i}$, one can see that every
$\u{\nu}\in {\Bbb Z}^2\setminus\{ \u{0}\}$ goes into $I^-$ after 
sufficiently
many iterations of matrices $N_{a_i}$ (as the eigenvector
of $N_{a_i}^{-1}$ with an eigenvalue of norm larger than 1 points into $I^-$).
In other terms, a resonant mode at some scale turns out to be
nonresonant at a sufficiently smaller scale. We notice that $\u{0}$
is not an element of $I^-$, i.e., it is resonant.\\
We eliminate completely all the nonresonant modes of the perturbation by a 
canonical transformation, connected to the identity, which is defined
by iterating KAM-type transformations (one iteration reduces the nonresonant
modes of the perturbation from $\varepsilon$ to $\varepsilon^2$).\\
One iteration step is performed by a Lie transformation  
$\mathcal{U}_S:
(\u{A},\u{\phi})\mapsto (\u{A}',\u{\phi}')$ generated by a function $S(\u{A},\u{\phi})$.
The expression of the Hamiltonian in the new coordinates is given by 
\begin{eqnarray}
H'&=& H\circ \mathcal{U}_S = e^{\hat{S}}H \nonumber \\
 &=& H+\{ S,H\} +\frac{1}{2!}\{S,\{ S,H\}\}+\cdots, \label{eqn:Hnc}
\end{eqnarray}
where $\{\, , \, \}$ is the Poisson bracket of two functions of the 
action and angle coordinates
$$
\{f,g\}=\frac{\partial f}{\partial \u{\varphi}}\cdot 
        \frac{\partial g}{\partial \u{A}}
       -\frac{\partial f}{\partial \u{A}}\cdot 
        \frac{\partial g}{\partial \u{\varphi}},
$$
and the operator $\hat{S}$ is defined as $\hat{S}H=\{S,H\}$. 
The generating function $S$ is chosen such that the order $\varepsilon$
of the nonresonant part (the modes in $I^-$) of the perturbation vanishes. We construct 
recursively the Hamiltonians $H_n$, starting with $H_1=H'$, such that the limit $H_{\infty}$
is canonically conjugated with $H'$ but does not contain nonresonant modes.
One step of this elimination procedure,
$H_n\mapsto H_{n+1}$, is done by applying a change 
of coordinates $\mathcal{U}_n$ such that the order of the nonresonant
modes of $H_{n+1}=H_n\circ\mathcal{U}_n$ is $\varepsilon_n^2$, where
$\varepsilon_n$ denotes the order of the nonresonant modes of $H_n$.
At the $n$-th step, the order of the nonresonant modes of $H_n$ is
$\varepsilon_0^{2^{n-1}}$, where $\varepsilon_0$ is the order of the nonresonant
modes of $H'$. If this procedure converges, it defines a canonical transformation
$\mathcal{U}_S=\mathcal{U}_1\circ\mathcal{U}_2\circ\cdots\circ\mathcal{U}_n
\circ\cdots$, such that the final Hamiltonian $H_{\infty}=H'\circ\mathcal{U}_S$
does not contain any nonresonant mode.\\
The specific implementation of this step can be performed in two versions : 
the first one is
a transformation acting in a space of
quadratic Hamiltonians in the actions of the form
\begin{equation}
\label{eqn:Hquad}
H({\u A},{\u \varphi})={\u \omega}_{0}\cdot{\u A}+
    \frac{1}{2}m({\u \varphi})( {\u \Omega}\cdot{\u A})^{2} 
       + g({\u \varphi}){\u \Omega}\cdot{\u A}
       + f({\u \varphi}),
\end{equation}
following
Thirring's approach~\cite{Bthir92,chan98b,chan98c}, where $m$, $g$, and $f$ are
three scalar functions of the angles. The second version
is a transformation acting on a 
more general family of Hamiltonians, with a power series expansion in the actions,
\begin{equation}
\label{EQN:HGENE}
   H(\u{A},\u{\varphi})=\u{\omega}_0\cdot\u{A}+\sum_{j=0}^\infty f^{(j)}(\u{\varphi})
   (\u{\Omega}\cdot\u{A})^j,
\end{equation}
following Ref.~\cite{abad98}, where $\{ f^{(j)}\}$ are scalar
functions of the angles. The key-point for the implementation for
quadratic Hamiltonians is that the KAM-transformations do not reduce the 
nonresonant modes of the quadratic part of the Hamiltonian. These transformations
are performed by Lie transformations generated by functions linear in the actions.
Following this procedure, the iterated Hamiltonians $H_n$ remain quadratic in
the actions.\\
The equations defining the elimination part for 
Hamiltonians (\ref{eqn:Hquad}), are given in Refs.~\cite{chan98b,chan99c}. 
Concerning those for Hamiltonians (\ref{EQN:HGENE}),
they are given in the Appendix.\\

The convergence of the elimination procedure (step 4, definition of $\mathcal{U}_S$)
has been rigorously
analyzed in Ref.~\cite{koch99} for the second version of the transformation.
 It has been proven that
for a sufficiently small perturbation, the elimination converges. Concerning the
quadratic case, we lack, at this moment, a theoretical framework to
prove an analogous theorem. The convergence of the elimination procedure 
outside the perturbative regime is
observed in both cases numerically.  

In summary, the RG-transformations for a given torus,
act as follows : First, some of the resonant modes of the perturbation
are turned into nonresonant modes by a frequency
shift and a rescaling in phase space. Then, a KAM-type iteration
eliminates these non-resonant modes, while producing some new
resonant modes.\\
Essentially, both versions of the RG-transformation give qualitatively and
quantitatively the same results for the class of invariant tori considered in this
article. The main advantage of the first version is that it is numerically 
more efficient as one only works with three scalar functions of the angles.
But from a theoretical point of view, there are some advantages to work with the second
version, since the first one leads asymptotically to non-analytic Hamiltonians 
for the attracting fixed sets (in the quadratic
part in the actions)~\cite{chan98b}. 

\section{Numerical results}
\label{sec:res}
 
For a given frequency $\omega$ (with bounded entries in its continued fraction expansion),
the numerical implementation of the 
RG-transformation shows that there are
two main domains in the space of Hamiltonians~: one where the iteration
converges towards a family of integrable Hamiltonians (trivial fixed set),
and the other where it diverges to infinity. These domains are separated by a surface
called {\em critical surface} (which is conjectured to coincide with the set of 
Hamiltonians which have a non-smooth critical
invariant torus of the considered frequency) and denoted $\mathcal{S}(\omega)$ 
in what follows.\\
The domain of attraction of the trivial fixed set is the domain where the
perturbation of the iterated Hamiltonians tends to zero. However, the
renormalization trajectories do not, in general, converge to a fixed Hamiltonian but to
a trajectory related to the Gauss map (\ref{eqn:gauss}).
The trivial fixed set is composed of Hamiltonians of the form
\begin{equation}
\label{eqn:Htri}
H_l(\u{A})=\u{\omega}_l\cdot \u{A}+\frac{1}{2}(\u{\Omega}_l\cdot\u{A})^2,
\end{equation}
where $\u{\omega}_l=(\omega_l,-1)$ and $\u{\Omega}_l=(1,\alpha_l)$.
The renormalization map transforms the vectors $\u{\omega}_l$ and $\u{\Omega}_l$
following the Gauss map
\begin{eqnarray*}
&& \omega_{l+1}=\omega_l^{-1}-[\omega_l^{-1}],\\
&& \alpha_{l+1}=(\alpha_l+[\omega_l^{-1}])^{-1}.
\end{eqnarray*}
Thus the trivial fixed set can be a fixed point, a periodic cycle, or in general
a set of Hamiltonians labeled by a trajectory of the Gauss map, i.e., by the 
asymptotic entries of the continued fraction expansion of $\omega$.\\
Correspondingly, on the critical surface $\mathcal{S}(\omega)$, 
the renormalization flow converges to a periodic cycle
if the frequency $\omega$ is a quadratic irrational (its continued fraction
expansion is asymptotically periodic), and
to a low
dimensional attracting set which is a strange chaotic attractor, in the case where
$\omega$ is not quadratic. The nontrivial attractor has a codimension 1 
stable manifold, i.e.,
one expansive direction transverse to the critical surface $\mathcal{S}(\omega)$. \\
We take the definitions as formulated by Grebogi {\em et al.}~\cite{greb84} : An attractor
of a map is called strange if it is not a finite number of points, nor a
piecewise differentiable set. An attractor is chaotic if typical orbits 
on it have a positive Lyapunov exponent.\\
 
{\em Quadratic irrational frequencies}-- We start by analyzing
the effect on $\o$ and $\alpha$, of $s$ renormalization steps.
Denote by $\{b_j\}$ the continued fraction expansion of $\alpha$ :
$\alpha=[b_0,b_1,\ldots]$. The renormalization $\mathcal{R}_{a_{s-1}}
\mathcal{R}_{a_{s-2}}\cdots\mathcal{R}_{a_0}$ changes $\o=[a_0,a_1,\ldots]$
into $[a_s,a_{s+1},\ldots]$, and $\alpha$ into $[a_{s-1},a_{s-2},\ldots,a_0,
b_0,b_1,\ldots]$.\\
If $\o$ has a periodic continued fraction expansion
of period $s>1$, i.e., $\o=[(a_1,\ldots,a_s)_\infty]$, we notice that
$\alpha$ converges to $[(a_s,\ldots,a_1)_\infty]$.
Therefore $\u{\Omega}$ converges to the unstable eigenvector of the
matrix $N_{a_s}\cdots N_{a_1}$ (the stable eigenvector of this matrix is
$\u{\o}_0$). \\
In that case, the RG-transformation has no fixed 
points but two periodic cycles with period $s$ : a trivial cycle which is attractive in
all the directions in the space of Hamiltonians, and a hyperbolic (critical) cycle with a 
codimension 1 stable manifold. The trivial cycle characterizes the domain of Hamiltonians 
that have a smooth invariant torus of the considered frequency, and the nontrivial cycle,
the domain where the torus is critical.\\
These
periodic cycles can be equivalently considered as fixed points of the 
RG-transformation defined by the composed operator
$\mathcal{R}_{s}=\mathcal{R}_{a_s}\mathcal{R}_{a_{s-1}}\cdots\mathcal{R}_{a_1}$.\\
The nontrivial fixed point of $\mathcal{R}_{s}$ associated
with $\o$ defines a universality class that we characterize with 
critical exponents such as the total rescaling of phase space (product
of the $s$ rescalings), and the unstable eigenvalue of the linearized map
around the fixed point.\\
The interpretation of these attracting cycles (trivial and non-trivial) of the
RG-trajectories in terms of the structure in phase space, of the invariant tori is the
following~:
In the perturbative regime, there exists a geometrical accumulation 
of a sequence of periodic orbits.
The fact that it also happens in the
critical case, but with a nontrivial ratio, implies universal self-similar
properties of the critical torus~\cite{kada81,shen82}.\\
Two frequencies $\o_1$ and $\o_2$ having the same periodic tail
(and different first entries) in their continued fraction expansions ($\o_1$ and
$\o_2$ are called equivalent),
belong to the same universality class : After a finite number of steps of the RG-transformation,
the equations defining the renormalization trajectories for $\o_1$ are the same as
those for $\o_2$. Thus they have the same critical exponents and scaling factors.
The initial integers in the continued
fraction expansion are irrelevant for the breakup of the torus.\\
Associated with this nontrivial fixed point, we also have  nontrivial
fixed sets related to the nontrivial fixed point by 
symmetries~\cite{mack95,koch99,chan98a,chan99c}, which therefore 
belong to the same universality class.  \\

\begin{figure}
\begin{center} 
\unitlength 1cm
\centerline{
\begin{picture}(8,8)
\put(0,0.25){\psfig{figure=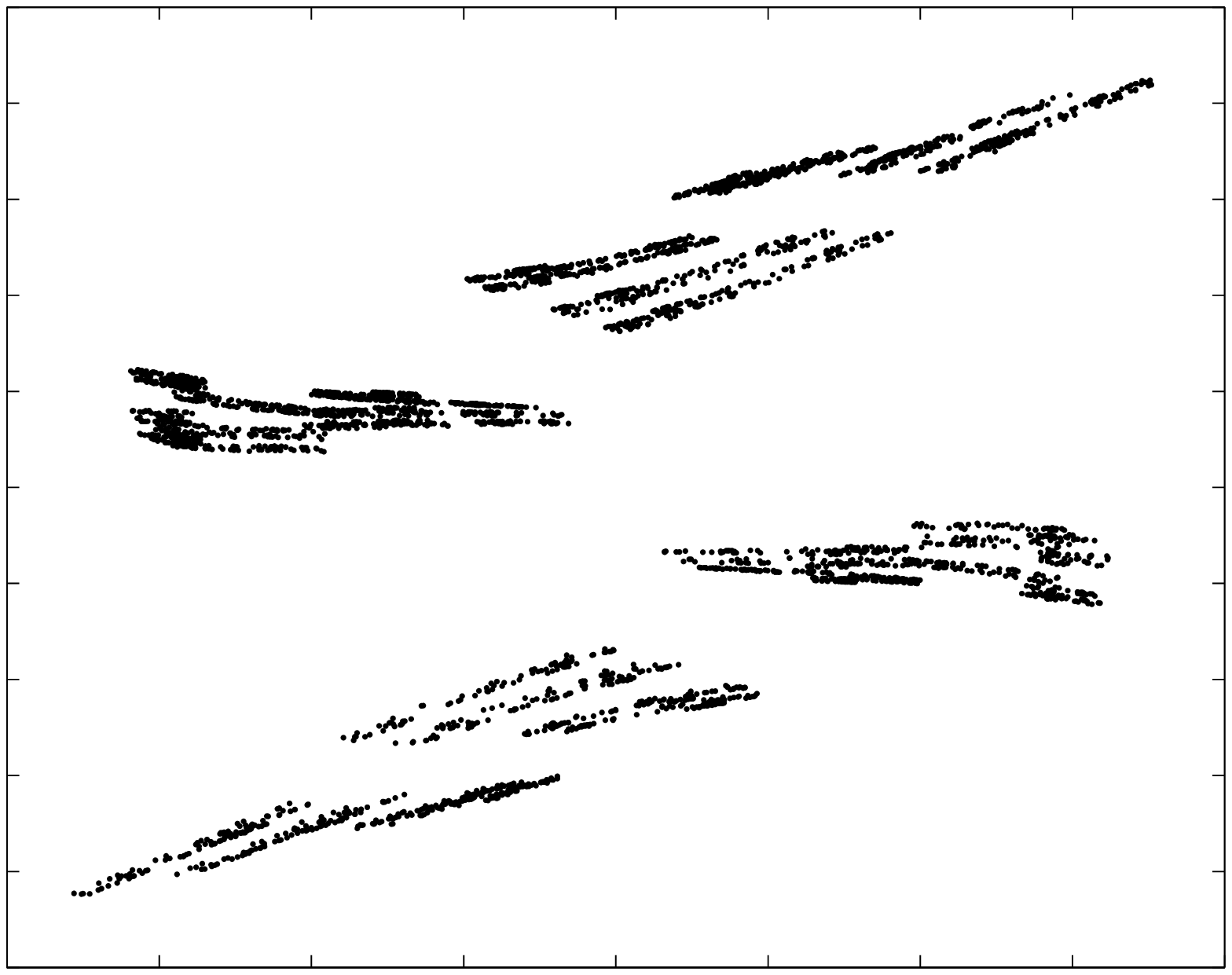,width=7cm,height=7cm}}
\put(3.55,0.35){$f_{\u{\nu}_1}^{(1)}$}
\put(-0.1,3.85){$f_{\u{\nu}_1}^{(2)}$}
\put(0,0.5){$-8\times 10^{-3}$}
\put(6.5,0.5){$8\times 10^{-3}$}
\put(-0.45,0.75){$-10^{-2}$}
\put(-0.2,7){$10^{-2}$}
\end{picture}}
\end{center}
\caption{\label{fig:SA}Projection on the plane $(f_{\u{\nu}_1}^{(1)}, f_{\u{\nu}_1}^{(2)})$,
of the critical attractor for a frequency 
whose continued fraction expansion is a random sequence of 1 and 2 with probability 
$P(1)=1/2$.}
\end{figure}

\begin{figure}
\begin{center} 
\unitlength 1cm
\centerline{
\begin{picture}(8,8)
\put(0,0.25){\psfig{figure=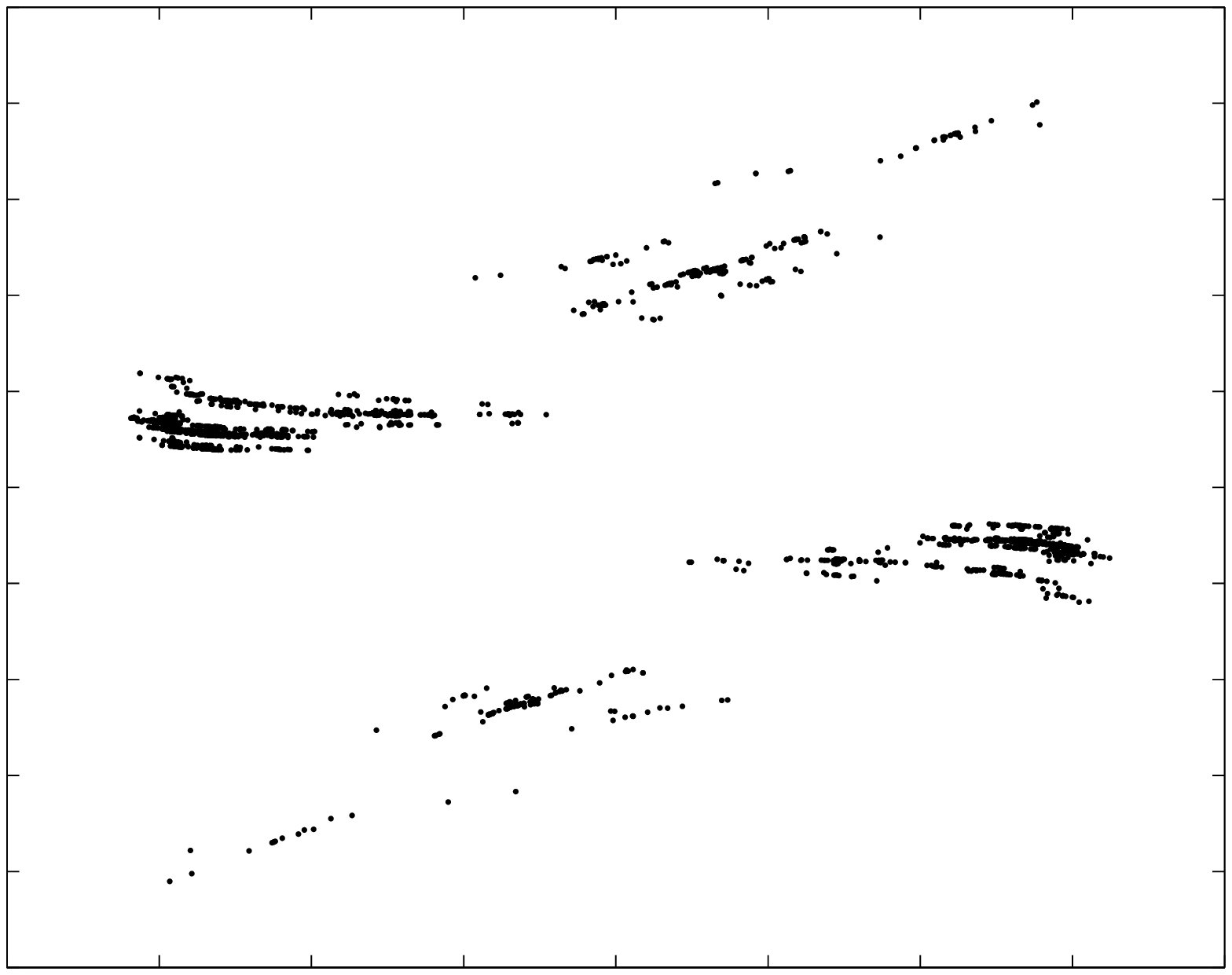,width=7cm,height=7cm}}
\put(3.55,0.35){$f_{\u{\nu}_1}^{(1)}$}
\put(-0.1,3.85){$f_{\u{\nu}_1}^{(2)}$}
\put(0,0.5){$-8\times 10^{-3}$}
\put(6.5,0.5){$8\times 10^{-3}$}
\put(-0.45,0.75){$-10^{-2}$}
\put(-0.2,7){$10^{-2}$}
\end{picture}}
\end{center}
\caption{\label{fig:SA9}Projection on the plane $(f_{\u{\nu}_1}^{(1)}, f_{\u{\nu}_1}^{(2)})$,
of the critical attractor for a frequency 
whose continued fraction expansion is a random sequence of 1 and 2 with 
probability $P(1)=9/10$.}
\end{figure}

{\em Nonquadratic~irrational~frequencies}-- The RG-transformations constructed 
in Sec.~\ref{sec:RG} are in principle defined for an arbitrary irrational 
frequency $\o\in]0,1[$.
They are based on the continued fraction expansion of the frequency. However, the numerical
analysis of the transformation on the critical surface $\mathcal{S}(\o)$, requires
to have a large number of entries in the continued fraction of $\o$ (in order to
iterate the RG-transformation on $\mathcal{S}(\omega)$
a sufficient number of times so as to observe the convergence to
an attracting set).
We consider frequencies whose entries in the continued fraction expansion are chosen randomly
in a finite set of integers, for instance, 1 or 2. Each entry in the continued
fraction expansion is chosen independently of the others according to the probability
$P(1)=p$ and $P(2)=1-p$. For $p=1$, the frequency is equal to the golden mean $\gamma$, and  
it is equal to $\sqrt{2}-1$ for $p=0$.\\
For a given frequency $\o$ with $p\in[0,1]$, the numerical analysis shows that there
are two main attracting sets in the space of Hamiltonians : a trivial fixed set composed
by integrable Hamiltonians, and a nontrivial one which lies on the critical surface 
$\mathcal{S}(\o)$.\\
Figure~\ref{fig:SA} depicts a projection of the critical attractor obtained
for a typical frequency with $p=1/2$, and Fig.~\ref{fig:SA9} for a frequency with $p=9/10$, 
on the plane $(f_{\u{\nu}_1}^{(1)},
f_{\u{\nu}_1}^{(2)})$ where $f_{\u{\nu}_1}^{(i)}$ denotes the Fourier coefficient
$\u{\nu}_1$ of the function $f^{(i)}$ of a Hamiltonian (\ref{EQN:HGENE})
on the critical attractor. For numerical purposes, these pictures have been obtained by 
truncating the
Hamiltonians on which the RG-transformation acts, in power series (we neglect higher
orders than $(\u{\Omega}\cdot \u{A})^J$, with $J=5$), and in Fourier series of
the functions $f^{(j)}$ (we neglect the Fourier modes with $\max_i|\nu_i|> L$,
with $L=5$).\\
 
\begin{figure}
\begin{center} 
\unitlength 1cm
\centerline{
\begin{picture}(8,8)
\put(0,0.25){\psfig{figure=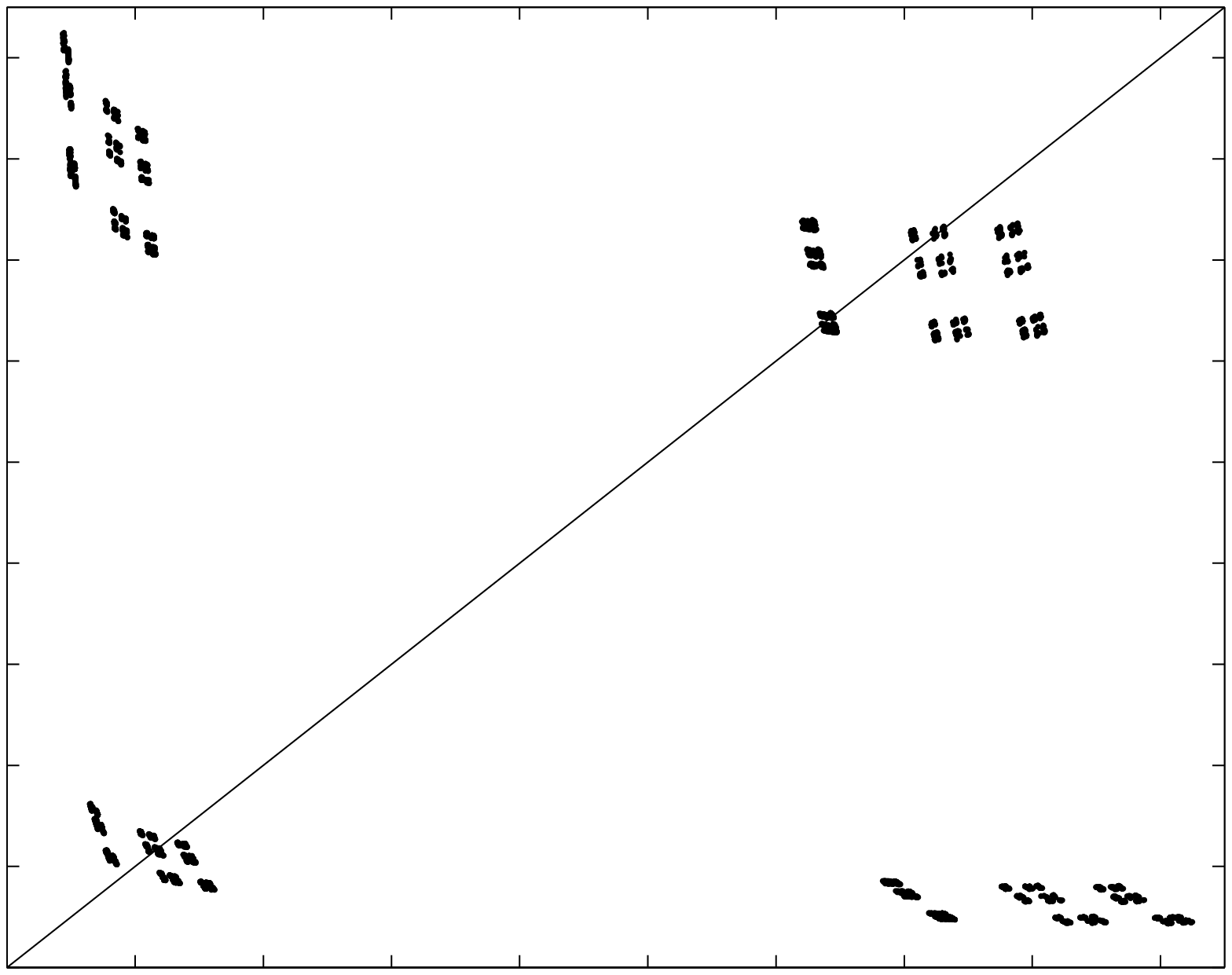,width=7cm,height=7cm}}
\put(3.45,0.25){$\lambda_j$}
\put(-0.5,3.75){$\lambda_{j+1}$}
\put(0.2,0.3){$2$}
\put(6.45,0.3){$20$}
\put(0,0.5){$2$}
\put(-0.2,6.65){$20$}
\end{picture}}
\end{center}
\caption{\label{fig:rescaling}Values of the rescalings $(\lambda_i,\lambda_{i+1})$
after $i$ iterations on the critical attractor of Fig.~\ref{fig:SA}.}
\end{figure}

A RG-trajectory on the critical attractor displays different scaling factors and 
critical exponents 
at each point of the trajectory. The distribution of these exponents depends only on the
frequency of the torus.\\
In Fig.~\ref{fig:rescaling}, we depict the values of the rescalings $(\lambda_{i},
\lambda_{i+1})$, where $\lambda_i$ denotes the value of the rescaling 
value~(\ref{eqn:rescaling})
after $i$ iterations on the critical attractor. This picture is obtained for a frequency
$\o$ with $p=1/2$, and the cut-off parameters are $J=5$ and $L=5$.
There are four main parts in this graph.
They correspond to the four possible changes of the first entry (to the second one) of the
continued fraction of the frequency : $1\rightarrow 1$, $1\rightarrow 2$,
$2\rightarrow 1$, $2\rightarrow 2$.
There are two values for which $\lambda_{i+1}=\lambda_i$ :
4.339 and 14.871 which are the values for the rescaling for the frequencies
$(\sqrt{5}-1)/2$ and $\sqrt{2}-1$~\cite{shen82,mack89b}.\\
 
If we consider two frequencies associated with a same probability $P(1)=p$, the 
numerically computed projections
on the Fourier modes of the perturbation,
of the critical attractors for each frequency, look identical.
This suggests that the properties of critical tori do not depend on the order of the
entries in the continued fraction expansion, but only on the distribution of the
values of the entries. In other words, critical tori with frequencies $\o_1$ and
$\o_2$ associated with a same probability distribution in the continued fraction
expansion,
are {\em statistically} self-similar. We notice that this discussion encompasses also the
case of two equivalent frequencies.\\

For two frequencies associated to different distributions of the entries in the continued
fraction expansion (for different values of $p$), the support of 
projections (on the Fourier modes
of the perturbation) of the critical attractors are the same, but
the renormalization trajectories visit the attractor with different densities.\\
For instance, concerning the projection of the attractor found for 
a frequency with probability $p=9/10$ in
Fig.~\ref{fig:SA9}, 
since $p\geq 1/2$ the renormalization trajectory is more concentrated 
around the fixed point (or around a periodic cycle related to this fixed point
by symmetry) obtained for the 
golden mean ($p=1$).\\

In order to describe a more precise image of the renormalization flow,
we consider an enlarged space where we add to the space of Hamiltonians a direction
corresponding to the frequency of the considered torus. The RG-transformations act in this
enlarged space consisting of couples $(H,\omega)$
(we iterate a renormalization operator for a given Hamiltonian $H$ and for a given
torus characterized by its frequency $\omega$). The numerical results lead to
the following conjecture :
The enlarged space is divided into two main parts : one
where the RG-iteration converges to a trivial fixed set, that consists
of couples $(H_l,\omega_l)$, where $H_l$ is integrable (of the form (\ref{eqn:Htri})). 
This trivial fixed set attracts all points $(H,\omega)$
such that $H$ has a smooth invariant torus with frequency $\omega$. 
The dynamics on it is determined by the Gauss map. The domain of
attraction of this trivial fixed set is bounded by a critical surface $\mathcal{S}$
which is the set of the critical surfaces 
$\mathcal{S}(\omega)$, i.e., composed by points $(H,\omega)$ such that $H$ has a 
non-smooth invariant torus with frequency $\omega$. 
The surface $\mathcal{S}$ has a fractal structure~: 
In Sec.~\ref{sec:frac}, we display a section of this surface, and analyze the fractal diagram
of a given one-parameter family. On the critical surface $\mathcal{S}$,
there is a {\em single} critical attractor to which all
Hamiltonians on $\mathcal{S}$ are attracted. 
On these attractors (trivial or critical), we have subsets that are periodic cycles 
of all the
periods; these cycles correspond to quadratic irrational frequencies. 
For instance, we have two fixed points on the nontrivial attractor : one corresponding to the
golden mean $\gamma$, and the other one, to $\sqrt{2}-1$.
We also have on the critical attractor, cycles (periodic and nonperiodic)
related to the previous ones by symmetries, and therefore have identical properties.
We can consider
them as artefacts of the definition of the RG-transformation (for instance, when the
frequency is equal to the golden mean, the RG-transformation can be modified such that the
cycle of period three~\cite{chan98a,chan99c} becomes a fixed point).\\
The critical attractor is not 
irreducible~\cite{Bruel89a,Bruel89b}
because it contains all the fixed points, periodic cycles etc., corresponding to specific
frequencies (e.g., quadratic frequencies). For a typical frequency, we expect the RG-trajectory
to visit a subset which is dense in the whole attractor.\\

Two invariant tori with frequencies $\o_1$ and $\o_2$ belong to the same universality class
in the following sense : The properties of these tori at the threshold of their breakup
are given by the analysis of a single hyperbolic attractor of a given renormalization
transformation. These two tori display the same set of scaling factors and critical
exponents (eigenvalues of the linearized map at each point of the attractor), with
a different probability distribution, depending on the distribution of the entries 
in the continued fraction of the frequency. \\
Based on our numerical results, we can speak of a single generic universality class 
for the breakup
of invariant tori for Hamiltonian systems with two degrees of freedom. The universal 
properties are obtained from a single critical attractor of a RG-transformation.
The critical attractor itself is not universal as it depends on the specific implementation
of the transformation, but the properties derived from it, are universal.

\section{Fractal diagram}
\label{sec:frac}

\begin{figure}
\begin{center} 
\unitlength 0.5cm
\centerline{
\begin{picture}(16,16)
\put(0,0.5){\psfig{figure=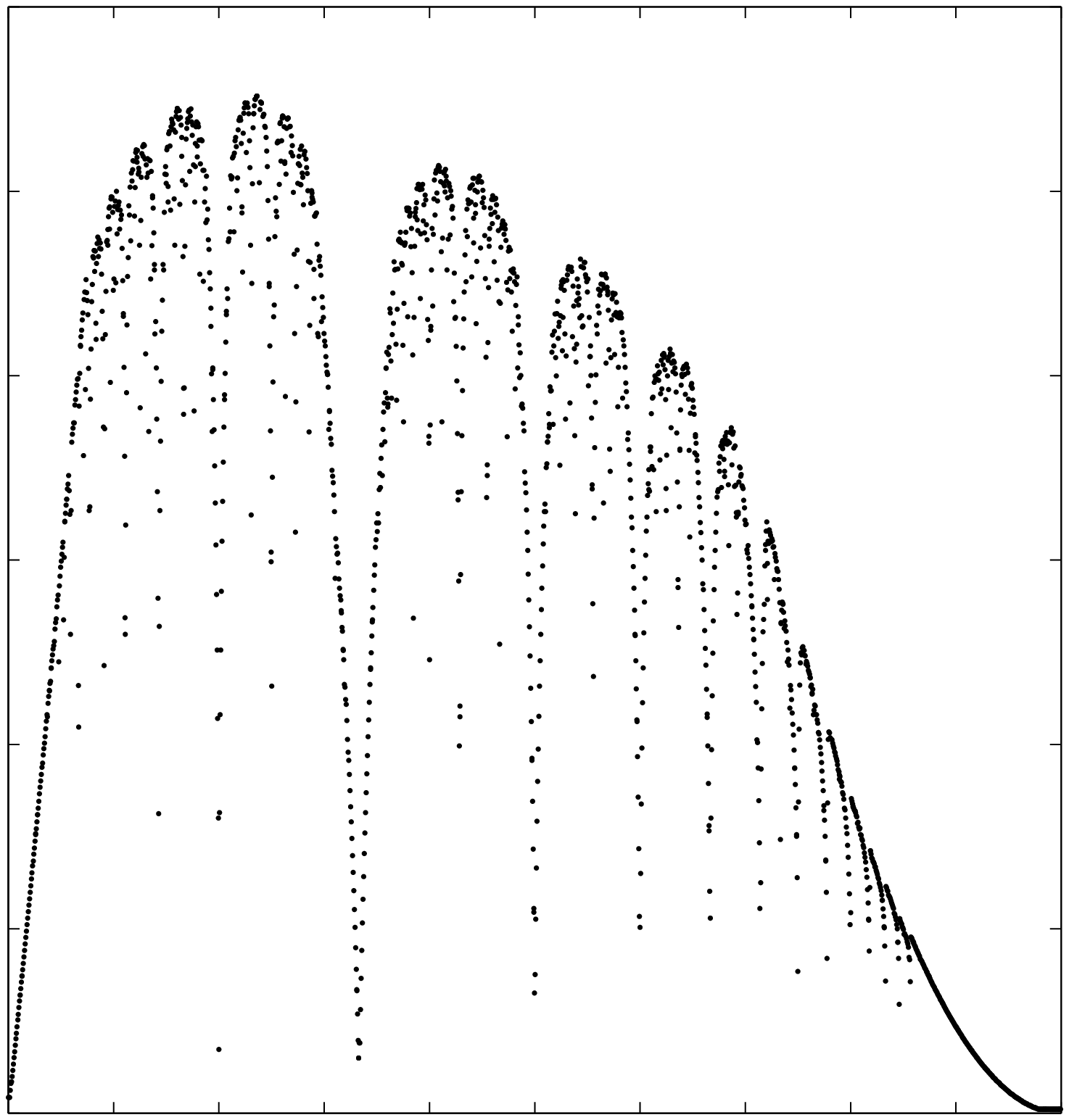,width=7cm,height=7cm}}
\put(7.2,-0.5){$\omega$}
\put(-0.6,7.6){$\varepsilon_c(\omega)$}
\put(0.87,0.6){$\uparrow$}
\put(7.27,0.6){$\uparrow$}
\put(3.45,0.6){$\uparrow$}
\put(5.15,0.6){$\uparrow$}
\put(8.54,0.6){$\uparrow$}
\put(13.65,0.6){$\uparrow$}
\put(3.89,13.3){$\downarrow$}
\put(3.89,13.9){$\gamma$}
\put(0.6,0.1){$1/2$}
\put(7,0.1){$3/4$}
\put(3.2,0.1){$3/5$}
\put(4.9,0.1){$2/3$}
\put(8.24,0.1){$4/5$}
\put(13.6,0.1){$1$}
\put(0.4,1){$0$}
\put(-0.2,14){$0.03$}
\end{picture}}
\end{center}
\caption{\label{fig:frac}Fractal diagram $\varepsilon_c(\omega)$
for the one-parameter family of Hamiltonians
(\ref{eqn:Hpara}).}
\end{figure}

In order to visualize the critical surface $\mathcal{S}$ in the enlarged space, in which the
critical attractor is contained, we represent a section of $\mathcal{S}$ by computing the 
fractal
diagram $\varepsilon_c(\omega)$ of the following one-parameter family of Hamiltonians :
\begin{eqnarray}
H_{\varepsilon}(\u{A},\u{\varphi})=&&\u{\o}_0\cdot \u{A}
+\frac{1}{2}(\u{\Omega}\cdot\u{A})^2\nonumber \\
&&+\varepsilon \left( \cos\varphi_1 +\cos(\varphi_1+\varphi_2)\right),\label{eqn:Hpara}
\end{eqnarray}
where $\u{\Omega}=(1,0)$ and $\u{\o}_0=(\omega,-1)$, where $\omega$ is the frequency
of the invariant torus. \\
For the golden mean $\gamma$, we checked that $\varepsilon_c(\gamma)$ coincides with
the value of the coupling where the RG-iteration starts to diverge, i.e., as
$n$ tends to infinity, we have
\begin{eqnarray}
\label{eqn:def1}
&& \mathcal{R}^nH_{\varepsilon} \to \infty \quad \mbox{ for }
\varepsilon>\varepsilon_c(\gamma),\\
&&\mathcal{R}^nH_{\varepsilon} \to H_0 \quad \mbox{ for }
\varepsilon<\varepsilon_c(\gamma). \label{eqn:def2}
\end{eqnarray}
We compute $\varepsilon_c(\omega)$, defined by Eqs.~(\ref{eqn:def1})-(\ref{eqn:def2}),
with $\mathcal{R}$ the renormalization transformation constructed for quadratic Hamiltonians
(\ref{eqn:Hquad}). We base our analysis on the conjecture that this $\varepsilon_c(\omega)$
coincides with the threshold at which the torus with frequency $\omega$ breaks up.
Each frequency $\omega$ of the diagram is chosen at random  
in the interval $[0.5,1]$. In order to iterate $\mathcal{R}$ (defined by a sequence of 
operators $\mathcal{R}_{a_i}$), we compute the ten first entries in the continued fraction 
expansion of $\omega$, and the tail of this expansion is filled with a sequence of 1. The
cut-off parameter for $\mathcal{R}$ (the restriction of the Fourier series to the modes
$\u{\nu}$ such that $\max_i |\nu_i|\le L$) is $L=10$.\\
The curve $\varepsilon_c(\omega)$ depicted
on Fig.~\ref{fig:frac}, indicates the fractal structure of the critical surface 
$\mathcal{S}=\bigcup_{\omega}\mathcal{S}(\omega)$. For rational frequencies, $\varepsilon_c$ 
is zero. In Fig.~\ref{fig:frac}, we observe strong resonances (Arnold tongues) close to
rational frequencies $\omega=p/q$ with small $q$. Similar pictures have been obtained
for the standard map~\cite{schm82,gree86,mack92b,marm92,star92}.\\
The curve $\varepsilon_c(\omega)$ for $\o\in [0,0.5[$ is obtained from the one for
$\o\in [0.5,1]$
by the symmetry $\varepsilon_c(\omega)=\varepsilon_c(1-\omega)$ which can be found by 
applying the canonical transformation 
$$
(\u{A},\u{\phi})\mapsto (\u{A}',\u{\phi}')=
(\tilde{T}\u{A},T\u{\phi}),
$$
where $\tilde{T}$ denotes the transposed matrix of
$$ T=\left( \begin{array}{cc} -1 & -1 \\ 0 & 1 \end{array} \right).$$
The effect of this transformation on $H_{\varepsilon}$ is just to change the
frequency from $\omega$ to $1-\omega$.\\
From Fig.~\ref{fig:frac}, we can see a numerical indication that the golden mean torus 
(with frequency 
$\gamma\approx 0.618$)
is the most robust one for the one-parameter family (\ref{eqn:Hpara}) : 
$\varepsilon_c(\gamma)>\varepsilon_c(\omega)$ for all other $\omega\in[0.5,1]$.\\

For each line going through the origin in the $MP$-parameter plane for Hamiltonians
\begin{eqnarray*}
H_{\varepsilon}(\u{A},\u{\varphi})=&&\u{\o}_0\cdot \u{A}
+\frac{1}{2}(\u{\Omega}\cdot\u{A})^2 \\
&&+M\cos\varphi_1 +P\cos(\varphi_1+\varphi_2),
\end{eqnarray*}
we can calculate in the same way, the critical coupling as a function of 
the frequency $\omega$. This
leads to pictures similar to Fig.~\ref{fig:frac}. Putting together all these sections,
we obtain a critical surface $\varepsilon_c(M,P,\omega)$ for this two-parameter family
of Hamiltonians and tori of frequencies $\omega$. The universal critical attractor
is contained in a surface of this kind $\varepsilon_c(\{ f_{\u{\nu}}^{(j)}\},\omega)$,
but with an infinite number of parameters $f_{\u{\nu}}^{(j)}$ which are the coordinates
of the considered space of Hamiltonians.

\section{A dimensional renormalization scheme}
\label{sec:simple}

In this section, we construct an approximate renormalization scheme based on
simple dimensional arguments. It is close to the type of transformations
considered by Escande and Doveil~\cite{esca81,esca85,mack88,chan99a}, and
by MacKay and Stark~\cite{mack89b}. This scheme is built by arguments based on the
dimensional analysis of the renormalization transformation.
The aim is to see
that the qualitative features of the renormalization flow obtained in the 
previous section, are already contained in a very simple scheme.\\

\begin{figure}
\begin{center} 
\unitlength 0.5cm
\centerline{
\begin{picture}(16,16)
\put(0,0.5){\psfig{figure=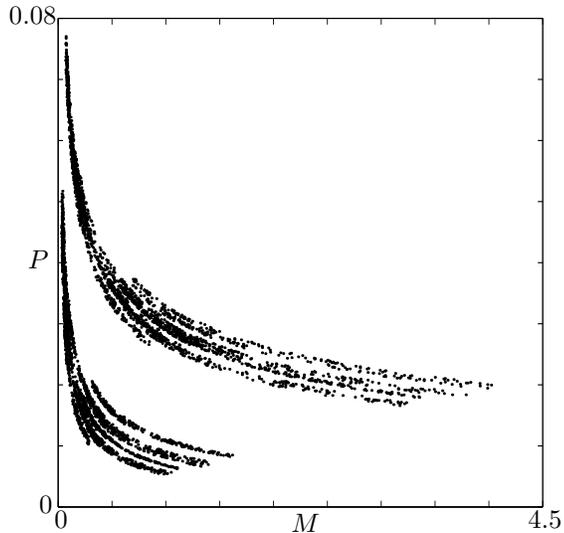,width=7cm,height=7cm}}
\put(7,0.5){$M$}
\put(0,7.5){$P$}
\put(0.7,0.6){$0$}
\put(13.3,0.6){$4.5$}
\put(0.3,1){$0$}
\put(-0.5,14){$0.08$}
\end{picture}}
\end{center}
\caption{\label{fig:SAapp}Critical attractor for the approximate renormalization
scheme for a frequency whose continued fraction expansion is a random sequence of 1 and 2
with probability $P(1)=1/2$.}
\end{figure}

We suppose
that the initial Hamiltonian contains only the two Fourier modes
$Me^{i\varphi_2}$ and $Pe^{i\varphi_1}$. The main scale is then 
determined by $\u{\nu}_{-1}=(0,1)$ and $\u{\nu}_{0}=(1,0)$. The next smaller
scale is represented by the mode $\u{\nu}_{0}=(1,0)$ together with the next 
resonance $\u{\nu}_{1}=(a_0,1)$, where $a_0$ denotes
the integer part of $\omega^{-1}$. The transformation is a change of 
coordinates that eliminates the mode $\u{\nu}_{-1}=(0,1)$, and produce the 
mode $\u{\nu}_{1}=(a_0,1)$. As $\u{\nu}_{1}=a_0\u{\nu}_{0}+\u{\nu}_{-1}$,
the amplitude of this mode is $MP^{a_0}$ (to the lowest order). 
Then we shift the Fourier modes~:
the mode $\u{\nu}_{1}$ (resp.~$\u{\nu}_{0}$) becomes the mode $\u{\nu}_{0}$
(resp.~$\u{\nu}_{-1}$). Consequently the frequency $\omega$ is changed into
$\omega'$ according to the Gauss map (\ref{eqn:gauss}).\\
We obtain thus the following RG-scheme :
\begin{eqnarray}
&& M'=k_1 P,\label{eqn:simple1}\\
&& P'=k_2 M P^{a_0}, \label{eqn:simple2}\\
&& \omega'=[\omega^{-1}]-a_0,
\end{eqnarray}
where $a_0=[\omega^{-1}]$.
In general, $k_i$ are functions of $\omega$, so the RG-scheme is equivalent
to a system 
(\ref{eqn:simple1})-(\ref{eqn:simple2}) driven by the Gauss map.\\ 
To give
an example, we consider $k_1=k_2=\omega^{-2}(a_0+\alpha)^2$, where $\alpha$
is determined by the inverse of the Gauss map $\alpha'=1/(a_0+\alpha)$. 
We have chosen these coefficients  
identical to the rescaling coefficient of the constant term in the actions for 
the renormalization explained in Sec.~\ref{sec:RG}. This choice is to a large extent
arbitrary, and is meant only as an illustration. The idea is that any function that
reflects the Gauss map should lead to qualitatively similar results for the attractor
and to the same critical exponents.\\ 
We denote $\mathcal{R}_{a_0}$ the following map, where $a_0=[\o^{-1}]$ :
$$
\mathcal{R}_{a_0}:(M,P,\o,\alpha)\mapsto (M',P',\o',\alpha').
$$
For a given frequency $\o=[a_1,a_2,\ldots]$, the transformation $\mathcal{R}$ is a sequence of 
$\mathcal{R}_{a_i}$. This transformation $\mathcal{R}$ has two main domains : one where
the iteration converges to $M=P=0$ (trivial fixed set), and one where the iteration diverges.
If the frequency is quadratic $\o=[b_1,b_2,\ldots,b_t,
(a_1,a_2,\ldots,a_s)_\infty]$, there is a hyperbolic (nontrivial) fixed point for
the transformation
$\mathcal{R}_{s}=\mathcal{R}_{a_s}\mathcal{R}_{a_{s-1}}\cdots\mathcal{R}_{a_1}$. 
The critical surface 
is the codimension 1 stable manifold of this nontrivial fixed point.\\
For nonquadratic irrational frequencies, the critical surface is the codimension 1 
stable manifold of a strange chaotic attractor : the chaoticity comes from the random
sequence of the $\mathcal{R}_{a_i}$.
We depict this critical attractor on Fig.~\ref{fig:SAapp}, computed using a frequency 
whose continued fraction expansion is a random sequence of 1 and 2 with $p=1/2$.
The largest Lyapunov exponent~\cite{bene76,bene80a,bene80b} $\kappa$ gives the
link between various Hamiltonians (of a given one-parameter family) near the critical
surface. It measures how far are two Hamiltonians $H_1$ et $H_2$ (near the critical surface), 
as we iterate renormalization $\mathcal{R}$ :
$$
\mathcal{R}^n H_1 - \mathcal{R}^n H_2 \approx e^{\kappa n} (H_1-H_2).
$$
We numerically find that the value given by this simple scheme, is in good agreement
with the one given by the complete RG-transformation, 
$\kappa \approx 0.68$ for $p=1/2$. This exponent depends on $p$ because it is computed 
for a given RG-trajectory which visits the different regions of the attractor with
some $p$-dependent distribution. For $p=1$ (golden mean), 
$\kappa\approx 0.49$ and for $p=0$, $\kappa\approx 0.89$.

\section*{acknowledgments}

We acknowledge
useful discussions with G.\ Benfatto, G.\ Gallavotti, H.\ Koch,
and R.S.\ MacKay.
Support from EC Contract No.\ ERBCHRXCT94-0460 for the project
``Stability and Universality in Classical Mechanics'' is acknowledged.

\appendix
\section*{Equations defining the KAM transformations for Hamiltonians~(\ref{EQN:HGENE})}

   In this appendix, we describe one step of the elimination procedure
   $H\mapsto H'$, defined for Hamiltonians (\ref{EQN:HGENE}),
   by considering that $f^{(j)}$
   depends on a small parameter $\varepsilon$, such that ${\Bbb I}^-f^{(j)}$
   is of order $O(\varepsilon)$, where ${\Bbb I}^-f^{(j)}$ denotes the nonresonant
   part of $f^{(j)}$, i.e.,
   $$
   {\Bbb I}^-f^{(j)}(\u{\varphi})=\sum_{\u{\nu}\in I^-} 
   f^{(j)}_{\u{\nu}} e^{i\u{\nu}\cdot\u{\varphi}},
   $$
   where $f_{\u{\nu}}^{(j)}$ denotes the Fourier coefficient of $f^{(j)}$
   with frequency vector $\u{\nu}$.
   We define $H_0$ as
   \begin{equation}
   H_0(\u{A})=\u{\omega}_0\cdot\u{A}+\langle f^{(2)}\rangle
   (\u{\Omega}\cdot\u{A})^2.
   \end{equation}
   In order to eliminate the nonresonant modes
   of $f^{(j)}$ to the first order in $\varepsilon$, we perform
   a Lie transformation $ \mathcal{U} : (\u{A},\u{\varphi})\mapsto
   (\u{A}',\u{\varphi}')$
   generated by a function $S$ of the form
   \begin{equation}
   S(\u{A},\u{\varphi})=i\sum_{j=0}^{\infty} Y^{(j)}(\u{\varphi})(\u{\Omega}
   \cdot\u{A})^j +a\u{\Omega}\cdot\u{\varphi}.
   \end{equation}
The expression of the Hamiltonian in the new coordinates is given by Eq.~(\ref{eqn:Hnc}).
   The first order in the perturbation of this Hamiltonian is $V+\{S,H_0\}$.
   Then $S$ is
   determined by the following condition
   \begin{equation}
   {\Bbb I}^-\{S,H_0\}+{\Bbb I}^-V=0.
   \end{equation}
   The constant $a$ eliminates the linear term in the 
   $(\u{\Omega}\cdot\u{A})$-variable, $\langle f^{(1)}
   \rangle$, by requiring that $\langle \{S,H_0\}\rangle +\langle f^{(1)}
   \rangle\u{\Omega}\cdot\u{A}=0$ :
   \begin{equation}
   a=-\frac{\langle f^{(1)}\rangle}{2\Omega^2\langle f^{(2)}\rangle},
   \end{equation}
   and $Y^{(j)}$ is determined by
   \begin{eqnarray}
   && i\u{\omega}_0\cdot\u{\partial} Y^{(0)}
   + {\Bbb I}^-f^{(0)}=const,\\
   && i\u{\omega}_0\cdot\u{\partial}Y^{(j)}+{\Bbb I}^-f^{(j)}
   +2i\langle f^{(2)}\rangle\u{\Omega}\cdot\u{\partial} 
   Y^{(j-1)}=0, 
   \end{eqnarray}
   for $j\geq 1$, where $\u{\partial}$ denotes the derivative with respect to the angles~:
   $\u{\partial}\equiv\partial/\partial\u{\varphi}$. 
   These equations are solved by representing them in the Fourier
   space :
   \begin{eqnarray*}
    && Y^{(0)}(\u{\varphi})=\sum_{\u{\nu}\in I^-} \frac{f_{\u{\nu}}^{(0)}}{
       \u{\omega}_0\cdot\u{\nu}} e^{i\u{\nu}\cdot\u{\varphi}},\\
    && Y^{(j)}(\u{\varphi})=\sum_{\u{\nu}\in I^-} \frac{1}{
       \u{\omega}_0\cdot\u{\nu}}\left( f_{\u{\nu}}^{(j)}-
       2\langle f^{(2)}\rangle\u{\Omega}\cdot\u{\nu} Y_{\u{\nu}}^{(j-1)}
       \right)e^{i\u{\nu}\cdot\u{\varphi}},
   \end{eqnarray*}
   for $j\geq 1$. Then we compute $H'=
   H\circ\mathcal{U}$ be calculating recursively the Poisson brackets
   $\hat{S}^kH=\hat{S}\hat{S}^{k-1}H$, for $k\geq 1$. Denoting 
   $H_k=\hat{S}^kH$, $H'$ becomes
   \begin{equation}
   H'=\sum_{k=0}^{\infty}\frac{H_k}{k!}.
   \end{equation}
   We expand $H'$ in power series in the actions
   \begin{equation}
   H'(\u{A},\u{\varphi})=\u{\omega}_0\cdot\u{A}+\sum_{j=0}^{\infty} 
   f'^{(j)}(\u{\varphi})
   (\u{\Omega}\cdot\u{A})^j.
   \end{equation}
   The Hamiltonian $H'$ is expressed by the image of the functions $f^{(j)}$
   given by the following expressions
   \begin{equation}
   \label{eqn:series}
   f'^{(j)}=\sum_{k=0}^\infty \frac{f_k^{(j)}}{k!},
   \end{equation}
   where
   \begin{eqnarray*}
   && f_0^{(j)}=f^{(j)},\\
   && f_1^{(1)}=i\sum_{l=0}^j (j+1-l)\left(f^{(j+1-l)}\u{\Omega}\cdot
   \u{\partial} Y^{(l)}-Y^{(j+1-l)}\u{\Omega}\cdot\u{\partial}f^{(l)}\right) 
    +a\Omega^2(j+1)f^{(j+1)}-{\Bbb I}^- f^{(j)},\\
   && f_{k+1}^{(j)}=i\sum_{l=0}^j (j+1-l)\left(f_k^{(j+1-l)}\u{\Omega}\cdot
   \u{\partial} Y^{(l)}-Y^{(j+1-l)}\u{\Omega}\cdot\u{\partial}f_k^{(l)}\right) 
    +a\Omega^2(j+1)f_k^{(j+1)},
   \end{eqnarray*}
   for $k\geq 1$ and $j\geq 0$. Numerically, we compute $f'^{(j)}$ for
$j=0,1,\ldots,J$, by truncating
   the series (\ref{eqn:series}) to a finite sum over $0\leq k\leq K$. For the
   calculation of $f'^{(j)}_k$, it is not necessary to compute it
   for $j$ too large as its contribution in $H'$ might exceed the truncation 
   in the actions. More precisely, we compute $f'^{(j)}_k$ for 
   $j=0,\ldots,J+k(J-1)$ if $k$ is lower than $k_0=[K/J]+1$ (where $[x]$ denotes the integer
   part of $x$), and for $j=0,\ldots,J+K-k$ if $k\geq k_0$. For instance, if
   we truncate at $J=3$, we compute $\hat{S}H$ up to order $(\u{\Omega}\cdot
   \u{A})^5$.

\newpage

\end{document}